\documentclass[11pt]{article} 

\usepackage[utf8]{inputenc} 

\usepackage{geometry} 
\usepackage{graphicx} 

\usepackage{booktabs} 
\usepackage{array} 
\usepackage{paralist} 
\usepackage{verbatim} 
\usepackage{subfig} 
\usepackage[english]{babel}
\usepackage[utf8]{inputenc}
\usepackage[colorinlistoftodos, color=green!40, prependcaption]{todonotes}
\usepackage[pdftex, pdftitle={Article}, pdfauthor={Author}]{hyperref} 
\usepackage{amsmath, amssymb}
\usepackage{upgreek}
\usepackage{verbatim}
\usepackage{hyperref}
\usepackage{multicol}
\newcommand{\lcdm}{$\Lambda$CDM}
\newcommand{\om}{\Omega_{m0}}
\newcommand{\obh}{\Omega_b h^2}

\usepackage{fancyhdr} 
\usepackage{sectsty}
\allsectionsfont{\sffamily\mdseries\upshape} 

\usepackage[nottoc,notlof,notlot]{tocbibind} 
\usepackage[titles,subfigure]{tocloft} 

\title{Evidence against Ryskin's model of cosmic acceleration}
\author{Joseph Ryan \\ Department of Physics, Kansas State University \\ 116 Cardwell Hall, Manhattan, KS 66506, USA \\ email: \href{mailto:jwryan@phys.ksu.edu}{jwryan@phys.ksu.edu}}

\begin{document}
\maketitle

\abstract{
In this paper I examine how well Ryskin's model of emergent cosmic acceleration fits several sets of cosmological observations. I find that while Ryskin's model is somewhat compatible with the standard model of cosmic acceleration (\lcdm) for low redshift ($z \lesssim 1$) measurements, its predictions diverge considerably from those of the standard model for measurements made at high redshift (for which $z \gtrsim 1$), and it is therefore not a compelling substitute for the standard model.}

\vspace{2mm}

keywords: Cosmology, General Relativity, Dark energy

\section{Introduction} \label{sec:Intro}
Observations show that the universe is currently undergoing an accelerated phase of expansion, which was preceded by a decelerated phase of expansion at $z \gtrsim 0.75$ (for reviews of the standard paradigm, see e.g. Refs. \cite{Peebles_Ratra, Ratra_Vogeley}; for a discussion of the deceleration-acceleration transition, see Ref. \cite{Farooq_Ranjeet_Crandall_Ratra_2017} and references therein). In the standard cosmological model, \lcdm, this acceleration is powered by a spatially homogeneous energy density in the form of a cosmological constant, $\Lambda$. Although the cosmological constant has successfully explained many observations to date (see e.g. Refs. \cite{planck2018_overview, planck2018}), an explanation of its origin in terms of fundamental physics remains elusive (see e.g. Refs. \cite{Martin, Straumann, Weinberg_1, Weinberg_2}). Many researchers have therefore attempted to construct models of cosmic acceleration that do not incorporate the cosmological constant, or any other form of dark energy (for a review of which, see e.g. Ref. \cite{modgrav}). One such model (not covered in Ref. \cite{modgrav}) is Gregory Ryskin's model of emergent cosmic acceleration (presented in Ref. \cite{Ryskin}). In this model, the observed acceleration of the universe is argued to emerge naturally as a consequence of applying a mean-field treatment to Einstein's gravitational field equations on cosmic scales. In this way, Ryskin claims to have arrived at an explanation of cosmic acceleration that does not require any fundamentally new physics. According to Ref. \cite{Ryskin}, Ryskin's model accurately fits the Hubble diagram built from SNe Ia data, but I will show in this paper that there are other data sets with which Ryskin's model is much less compatible. In addition to predicting a value of the Hubble constant ($H_0$) that is larger than the values obtained from the CMB and from local measurements (see Refs. \cite{planck2018} and \cite{riess2018}, respectively, for these measurements), Ryskin's model fails to predict the trend in high-redshift ($z \gtrsim 1$) Hubble parameter data when its predicted Hubble parameter curve is plotted together with these data.

Recently, another group found that Ryskin's model can not accurately describe structure formation (see Ref. \cite{Revisit_Ryskin}), while leaving open the possibility that other types of observations may be compatible with this model. This paper is complementary to, and independent of, the analysis presented in Ref. \cite{Revisit_Ryskin}; I will show that none of the data sets I have collected favor Ryskin's model over \lcdm, making it unlikely that Ryskin's model will be saved by future measurements.

In Sec. II I briefly describe Ryskin's model, in Sec. III I describe the data that I use, and in Sec. IV I present my results.

\section{Emergent cosmic acceleration} \label{sec:Theory}
The central claim of Ryskin's paper is that the standard gravitational field equations of General Relativity,
\begin{equation}
\label{eq:Einstein}
    R_{\mu \nu} - \frac{1}{2}Rg_{\mu \nu} = \kappa T_{\mu \nu},
\end{equation}
where $\kappa := \frac{8 \pi G}{c^4}$, which are well-tested on the scale of the solar system, must be modified when applied to cosmological scales. Ryskin contends, in Ref. \cite{Ryskin}, that moving from sub-cosmological scales (in which matter is distributed inhomogeneously) to cosmological scales (in which matter is distributed homogeneously) introduces emergent properties to the description of the Universe (such properties being ``emergent" in the sense that they are not apparent on sub-cosmological scales) and that the observed large-scale acceleration of the Universe may be one such emergent property. According to Ryskin's model, the emergence of cosmic acceleration is therefore analogous to the emergence of properties like temperature and pressure that result from averaging over the microscopic degrees of freedom of an ideal fluid, thereby moving from a length-scale regime in which kinetic theory is valid to a length-scale regime in which the fluid must be described with continuum hydrodynamics; in the same way, Ryskin contends, cosmic acceleration ``emerges" from Einstein's field equations when these are applied to cosmological scales.  Ryskin's model, therefore, purports to offer an explanation of the origin of cosmic acceleration that does not require the introduction of dark energy. In his paper, Ryskin introduces a mean-field tensor with components $\kappa\Phi_{\mu \nu}$ into the right-hand side of Eq. \ref{eq:Einstein}, so that the (large-scale) field equations read
\begin{equation}
\label{eq:Ryskin}
    R_{\mu \nu} - \frac{1}{2}Rg_{\mu \nu} = \kappa \left(T_{\mu \nu} + \Phi_{\mu \nu}\right).
\end{equation}
It can be shown (see Ref. \cite{Ryskin} for details) that when the standard gravitational field equations are modified in this way, and that if the Universe has flat spatial hypersurfaces and is dominated by non-relativistic matter, then the total rest energy density and pressure of the averaged, large-scale cosmic fluid are
\begin{equation}
\label{eq:rho}
    \rho = 4\rho_m,
\end{equation}
and
\begin{equation}
\label{eq:p}
    p = -3\rho_m,
\end{equation}
respectively, where $\rho_m$ and $p_m$ are the rest energy density and pressure of the non-relativistic matter. Energy conservation then implies that $\rho_m \propto a^{-3/4}$, from which $a(t) \propto t^{8/3}$ follows. Rykin's model therefore predicts that the Hubble parameter takes the simple form
\begin{equation}
\label{eq:RyskinH}
    H(z) = H_0\left(1 + z\right)^{3/8},
\end{equation}
where the definitions of the Hubble parameter $H(t) := \frac{\dot{a}\left(t\right)}{a\left(t\right)}$ and redshift $1 + z := \frac{a\left(0\right)}{a\left(t\right)}$ have been used. The Hubble parameter derived in Ref. \cite{Ryskin} accurately fits the Hubble diagram constructed from SNe Ia data, which Ryskin takes as evidence that his model may be able to explain the origin of cosmic acceleration of the Universe without invoking dark energy. My goal in this paper is to test Eq. \ref{eq:RyskinH} against several sets of observational data (containing measurements at higher redshifts than the SNe Ia measurements used in Ref. \cite{Ryskin}), to determine whether or not Ryskin's model can fit these data sets as well as it fits the currently available SNe Ia data.

\section{Analysis} \label{sec:Analysis}
\subsection{Data} \label{subsec:Data}
In this paper I use 31 measurements of the Hubble parameter $H(z)$, 11 distance measurements derived from baryon acoustic oscillation (``BAO") data, and 120 quasar (``QSO") angular size measurements. The $H(z)$ data are listed in Ref. \cite{Ryan_1}; see that paper for a description. The BAO data are listed in Ref. \cite{Ryan_2}. My method of analyzing these data is slightly different from the method employed in Ref. \cite{Ryan_2}; see below for a discussion. The QSO data are listed in Ref. \cite{Cao_et_al2017b}; see that paper and Ref. \cite{Ryan_2} for a description and discussion.

For the $H(z)$ data set, the measured quantity is $H(z)$ itself, namely the Hubble parameter as a function of the redshift $z$. For the BAO data, the measured quantities are a set of distances $D_H(z)$, $D_M(z)$, and $D_V(z)$ (defined and described in Refs. \cite{Hogg, Ryan_2, Ryan_1}) and $H(z)$, scaled by the value that the sound horizon $r_{\rm S}$ takes at the drag epoch. This latter quantity depends on the homogeneous part of the dimensionless matter density parameter ($\Omega_{m0}$), the Hubble constant ($H_0$), the dimensionless baryon density parameter ($\Omega_b h^2$), and the CMB temperature ($T_{\rm CMB}$), thereby making $r_{\rm S}$ a model-dependent quantity.\footnote{$h := H_0/(100 \hspace{1mm} {\rm km}^{-1} \hspace{1mm} {\rm Mpc}^{-1})$} In this paper I use the same fitting formula that was used in Ref. \cite{Ryan_2} to compute $r_{\rm S}$, and I use the same CMB temperature (from Ref. \cite{Fixsen}), but depending on the analysis method (see below) I marginalize over $\Omega_b h^2$ and/or $\Omega_{m0}$. In all other respects my treatment of the BAO data here is the same as the treatment described in Ref. \cite{Ryan_2}. Finally, the measured quantity in the QSO data set is $\theta(z) = l_{m}/D_{A}(z)$, where $l_m = 11.03 \pm 0.25$ pc is a characteristic linear scale\footnote{Specifically, $l_m$ is the radius at which the jets of the QSOs tend to become opaque (when observed at frequency $f \sim 2$ GHz; see Ref. \cite{Cao_et_al2017b}).} and $D_A(z)$, the angular size distance, can be computed from $H(z)$. See Refs. \cite{Hogg} or \cite{Ryan_2} for a definition of $D_A(z)$, and Ref. \cite{Cao_et_al2017b} for a discussion of the characteristic linear scale $l_m$.
\subsection{Methods} \label{subsec:Methods}
I have chosen to analyze Ryskin's model according to two methods. In the first method, I compute the value that $H_0$ takes when the likelihood function, defined by
\begin{equation}
    \mathcal{L}(H_0) = \int e^{\chi^2(H_0)/2}\uppi(p_n) dp_n
\end{equation}
is maximized, within each data set separately and in full combination. I also compute the minimum $\chi^2$ corresponding to the best-fitting $H_0$, for which
\begin{equation}
\label{eq:chi2min}
    \chi^2_{\rm min} = -2{\rm ln}\left(\mathcal{L}_{\rm max}\right)
\end{equation}
This is very similar to the methods employed in Refs. \cite{Ryan_2, Ryan_1}; see those papers for details regarding the form that the $\chi^2$ function takes when it is computed within each model and for each data set. The prior function, $\uppi(p_n)$, is necessary to deal with the nuisance parameters $\Omega_{m0}$ and $\Omega_b h^2$ that enter the analysis through the calculation of the sound horizon $r_{\rm S}$ (see above). This prior function has the form $\uppi\left(\om, \obh\right) = \uppi\left(\om\right)\uppi\left(\obh\right)$, where \begin{equation}
\label{eq:om_prior}
  \uppi\left(\om\right) =
    \begin{cases}
      1 & \text{if $0.10 < \om < 0.70$}\\
      0 & \text{otherwise},
    \end{cases}       
\end{equation}
and
\begin{equation}
  \uppi\left(\obh\right) =
    \begin{cases}
      1 & \text{if $0.01000 < \obh < 0.05000$}\\
      0 & \text{otherwise}.
    \end{cases}       
\end{equation}
The main difference between this first analysis method and the analyses of \cite{Ryan_2, Ryan_1} is that I do not compare the best-fitting value of $H_0$ or the minimum value of $\chi^2$ in Ryskin's model directly to any other models (although the best-fitting $H_0$ can be compared to the measurements of $H_0$ made by Refs. \cite{planck2018} and \cite{riess2018}; see below). For the $\chi^2$ function I simply compare the minimum value of $\chi^2$ to the number of degrees of freedom $\nu$ (defined below), and conclude that the fit to the data is poor if $\chi^2_{\rm min}/\nu >> 1$. Additionally, I split the BAO data into two subsets, called ``BAO1" (containing all BAO measurements) and ``BAO2" (which excludes the measurements at $z > 2$), respectively, because the $\chi^2$ function for the $H(z)$ + QSO + BAO1 data combination is so large that the corresponding likelihood function evaluates to zero, and so can't be plotted. This is telling, because it suggests that Ryskin's model can not fit the observational data at high redshift (see also the discussion in Sec. \ref{sec:Results}). The combined fit therefore uses the $H(z)$ + QSO + BAO2 data combination (see Table \ref{tab:BFP} and Fig. \ref{fig:L(H0)}).

In my second analysis method, I directly compare the quality of the fit obtained in Ryskin's model to the quality of the fit obtained with a simple flat \lcdm\ model to each data set, considered separately. For the $H(z)$ and QSO data this is quite simple: all that is necessary is to plot either the function
\begin{equation}
    H(z) = H_0\sqrt{\om(1 + z)^3 + 1 - \om}
\end{equation}
(for the $H(z)$ data), or the function
\begin{equation}
    \theta(z) = \frac{l_m}{D_{\rm A}(z)}
\end{equation}
(for the QSO data) that is predicted by Ryskin's model, together with the same functions as predicted by \lcdm, and see which predicted function better fits the overall trend in the data (see Figs. \ref{fig:QSO_Lcomp}-\ref{fig:Hz_vs_z}). For the BAO data this kind of direct curve fitting is difficult to do, because several of the measurements are correlated (meaning that they do not have independent uncertainties), and the set as a whole consists of measurements of different things. In order to compare Ryskin's model to \lcdm\ using these data, I therefore redid the analysis of Ref. \cite{Ryan_2} (in that paper the constraints from BAO alone were not presented), allowing $H_0$ and $\Omega_{m0}$ to vary freely, with $\Omega_b h^2 = 0.02225$ for both Ryskin's model and \lcdm\footnote{This is the value that $\obh$ takes in the \lcdm\ model, previously used in Ref. \cite{Ryan_2}, and originally computed from Planck 2015 TT + lowP + lensing CMB anisotropy data in Ref. \cite{Park_Ratra}. It is also possible to marginalize over $H_0$ and $\Omega_{m0}$ so as to obtain a best-fitting value of $\Omega_b h^2$ in Ryskin's model, and then use this value instead of $\Omega_b h^2 = 0.02225$ in the two-parameter fits. Doing this turns out to be rather uninformative, however, as the best-fitting values of $\Omega_{m0}$ and $H_0$ that one obtains in this case turn out to be nearly identical to those obtained using $\Omega_b h^2 = 0.02225$.}.

\section{Results} \label{sec:Results}
\begin{figure*}
    \includegraphics[scale=1]{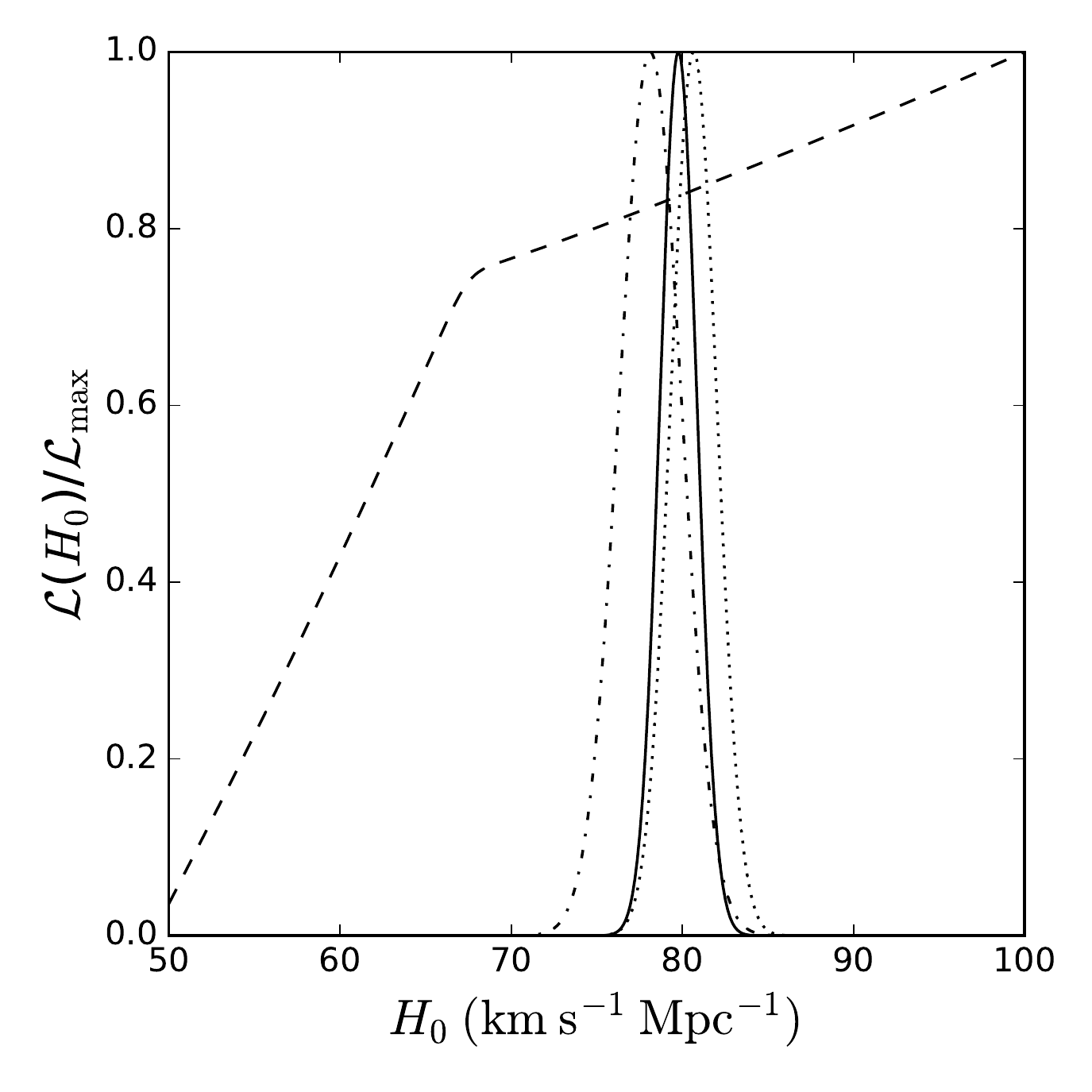}
    \caption{Likelihood functions for $H_0$ according to Ryskin's model. The dot-dashed curve represents the fit from the $H(z)$ data, the dotted curved represents the fit from the QSO data, the dashed curve represents the fit from the BAO2 data, and the solid curve represents the fit from the combined $H(z)$ + QSO + BAO2 data. See text for discussion.}
    \label{fig:L(H0)}
\end{figure*}

My results for the fit of Ryskin's model to the data are presented in Table \ref{tab:BFP} and Fig. \ref{fig:L(H0)}. In the first column of Table \ref{tab:BFP} I list the data combination, in the second column I list the one-dimensional best-fitting values of $H_0$ with their respective 1$\sigma$ and 2$\sigma$ uncertainties ($\sigma$ here being defined in the same way as the one-sided confidence limits used in Ref. \cite{Ryan_2}), and in the third column I list the corresponding value of $\chi^2_{\rm min}/\nu$, where $\chi^2_{\rm min}$ is computed from Eq. \ref{eq:chi2min}, and $\nu$ is the number of degrees of freedom:
\begin{equation}
\label{eq:nu}
    \nu = N - n - 1.
\end{equation}
In the above equation $N$ is the number of data points and $n$ is the number of model parameters.

\begin{table}
    \centering
    \caption{Best-fitting central values of $H_0$ (with 1 and 2$\sigma$ error bars) for the data combinations I considered.}
    \begin{tabular}{c|c|c}
        \hline
        Data set & $H_0$ (km s$^{-1}$ Mpc$^{-1}$) & $\chi^2_{\rm min}/\nu$\\
        \hline
        $H(z)$ & $78.12^{+1.82+3.64}_{-1.82-3.63}$ & 2.20\\
        QSO & 80.66$^{+1.35+2.70}_{-1.35-2.70}$ & 3.14\\
        BAO1 & $100_{-19.27-32.72}$ & 133.12\\
        BAO2 & $100_{-26.78-41.79}$ & 14.48\\
        $H(z)$ + QSO + BAO2 & $79.77^{+1.08+2.17}_{-1.08-2.17}$ & 3.38\\
        \hline
    \end{tabular}
    \label{tab:BFP}
\end{table}

From Table \ref{tab:BFP} and Fig. \ref{fig:L(H0)}, one can see that the best-fitting value of $H_0$ from the combination $H(z)$ + QSO + BAO2 (which gives the tightest error bars) is a little over 3$\sigma$ away from the measurement of $H_0 = 74.03 \pm 1.42$ km$^{-1}$ Mpc$^{-1}$ made by Ref. \cite{riess2018}, and over 10$\sigma$ away from the measurement of $H_0 = 67.4 \pm 0.5$ km$^{-1}$ Mpc$^{-1}$ made by Ref. \cite{planck2018} (here $\sigma$ is equal to $\sqrt{1.08^2 + \sigma_l^2}$ where $\sigma_l$ is the uncertainty of either of the two measurements given above). The agreement, therefore, between the predicted value of $H_0$ under Ryskin's model and the measurements from Refs. \cite{planck2018} and \cite{riess2018} is not very good. Further, the value of $\chi^2_{\rm min}/\nu$ ranges from 2.20 to 133.12 for the fit of Ryskin's model to each data set, and is equal to 3.38 for the $H(z)$ + QSO + BAO2 data combination. This suggests, independently of the comparison to the $H_0$ measurements made by Refs. \cite{planck2018} and \cite{riess2018}, that Ryskin's model does not provide a good fit to the data listed in Table \ref{tab:BFP}. In Fig. \ref{fig:L(H0)}, the dashed curve represents the likelihood function computed from BAO2, the dot-dashed curve represents the likelihood function computed from $H(z)$ data, the dotted curve represents the likelihood function computed from QSO data, and the solid curve represents the product of these likelihood functions. Here again one can see how far away the value of $H_0$ predicted by Ryskin's model is from the measurements made by Refs. \cite{planck2018} and \cite{riess2018} when $H_0$ is fitted to the $H(z)$ + QSO + BAO2 data combination.

\begin{figure*}
\begin{multicols}{2}
    \includegraphics[width=\linewidth]{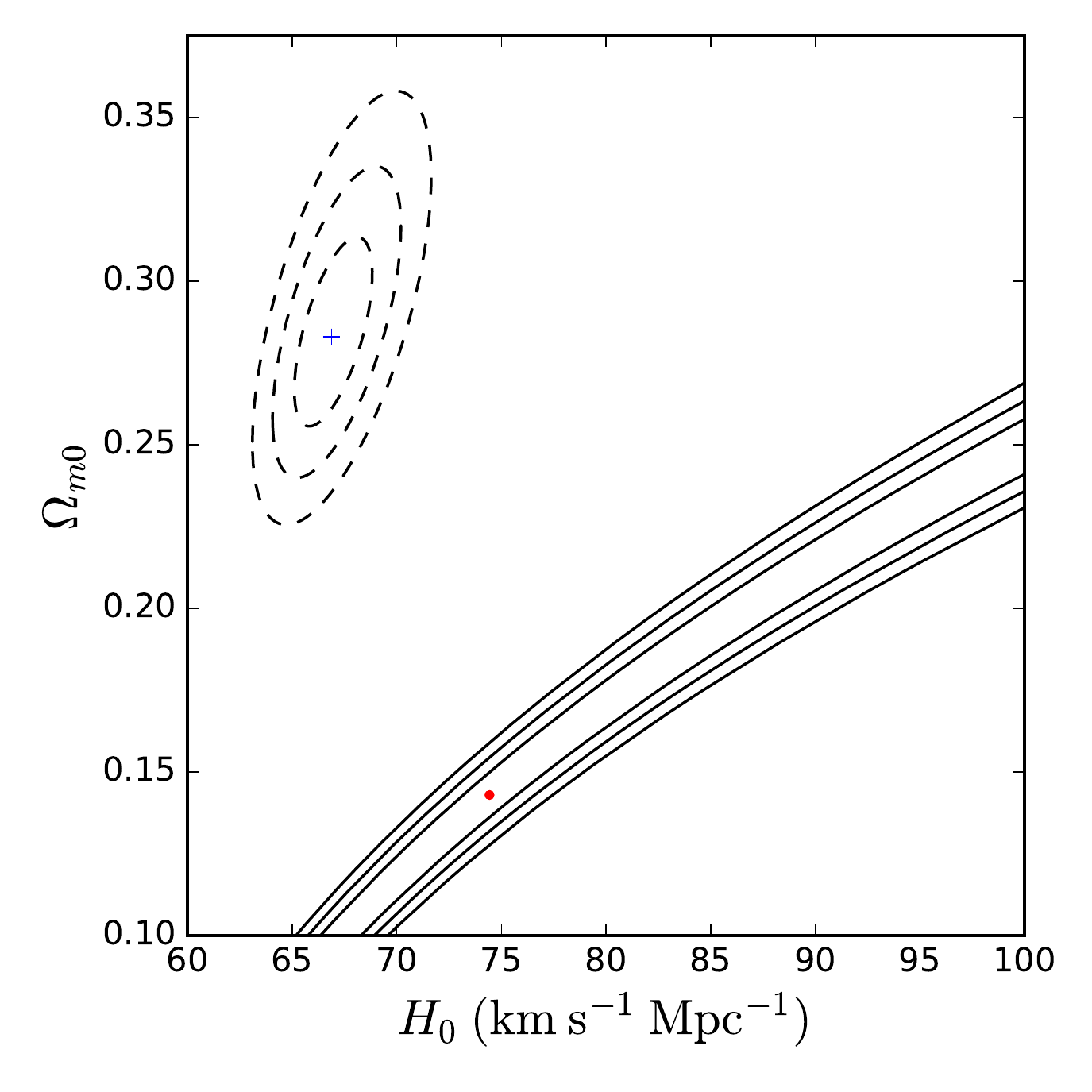}\par
    \includegraphics[width=\linewidth]{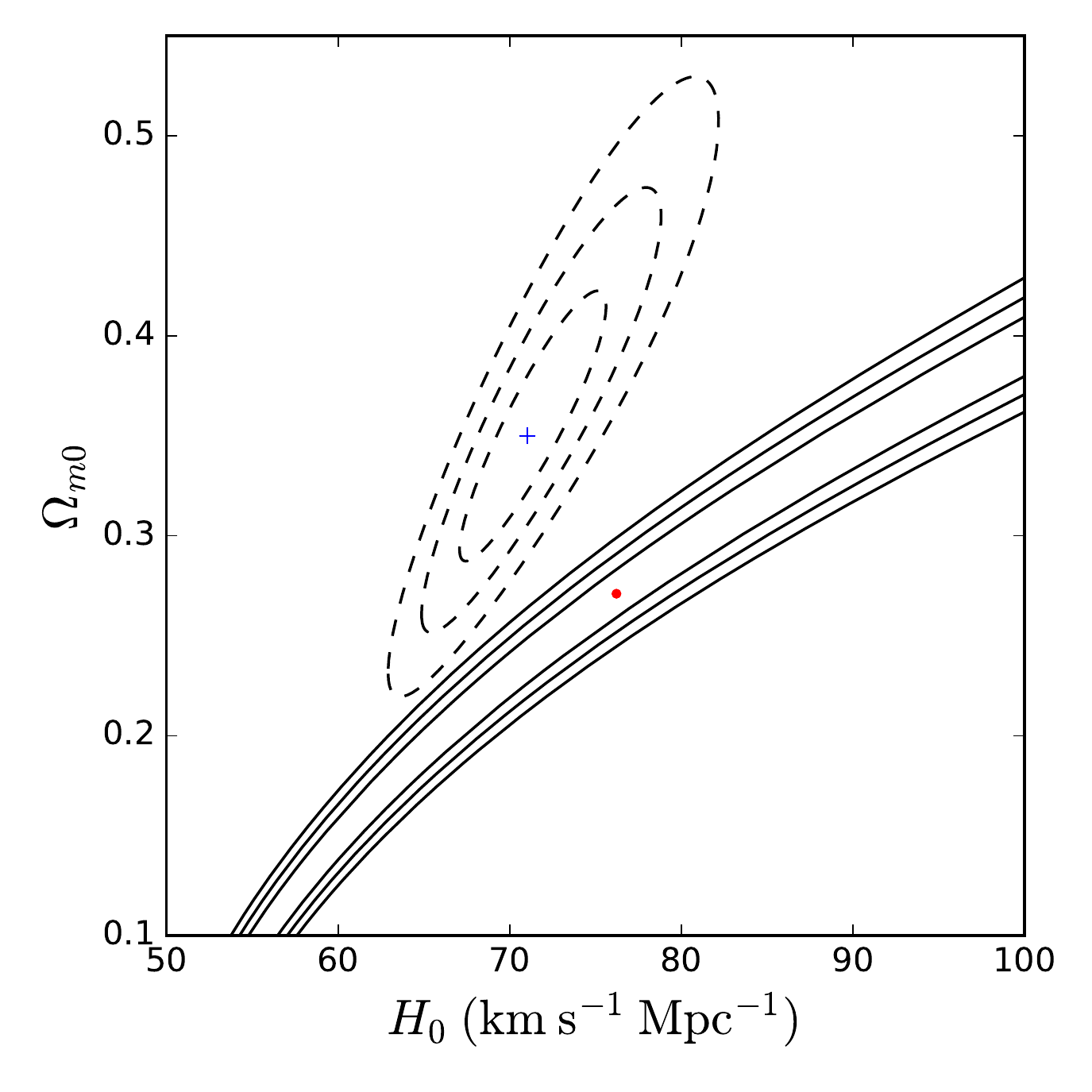}\par
\end{multicols}
\caption{Flat \lcdm\ model versus Ryskin's model with BAO data. The left panel corresponds to the BAO1 subset of the BAO data, and the right panel corresponds to the BAO2 subset of the same. In both columns I have plotted 1, 2, and 3$\sigma$ confidence contours and best-fitting points in $H_0$-$\Omega_{m0}$ space for both \lcdm\ and Ryskin's model. See text for discussion.}
\label{fig:Lcomp_2D}
\end{figure*}

\begin{table}
    \centering
    \caption{One- and two-dimensional best-fitting values of $H_0$ and $\Omega_{m0}$ for the BAO1 and BAO2 data combinations. Here $H_0$ has units of km s$^{-1}$ Mpc$^{-1}$ and $\chi^2_{\rm min}/\nu$ pertains to the two-dimensional fit.}
    \begin{tabular}{c|c|c|c|c|c}
        \hline
        Model & Data set & $H_0$ & $\Omega_{m0}$ & $\left(H_0, \Omega_{m0}\right)$ & $\chi^2_{\rm min}/\nu$\\
        \hline
        Ryskin & BAO1 & $100.0_{-21.43-31.06}$ & $0.237_{-0.0794-0.127}^{+0.00963+0.175}$ & (74.43, 0.143) & 147.87\\
         & BAO2 & $100.0_{-28.38-42.02}^{}$ & $0.373_{-0.145-0.249}^{+0.0167+0.0305}$ & (76.22, 0.271) & 14.56\\
        \lcdm & BAO1 & $66.91^{+1.322+2.715}_{-1.152-2.254}$ & $0.284^{+0.0205+0.0424}_{-0.0179-0.0348}$ & (66.88, 0.283) & 0.954\\
         & BAO2 & $71.28_{-2.593-5.003}^{+3.100+6.375}$ & $0.354^{+0.0491+0.102}_{-0.0409-0.0793}$ & (71.03, 0.350) & 0.650\\
        \hline
    \end{tabular}
    \label{tab:BFP_2D_1D_Lcomp}
\end{table}

\begin{figure*}
    \centering
    \includegraphics{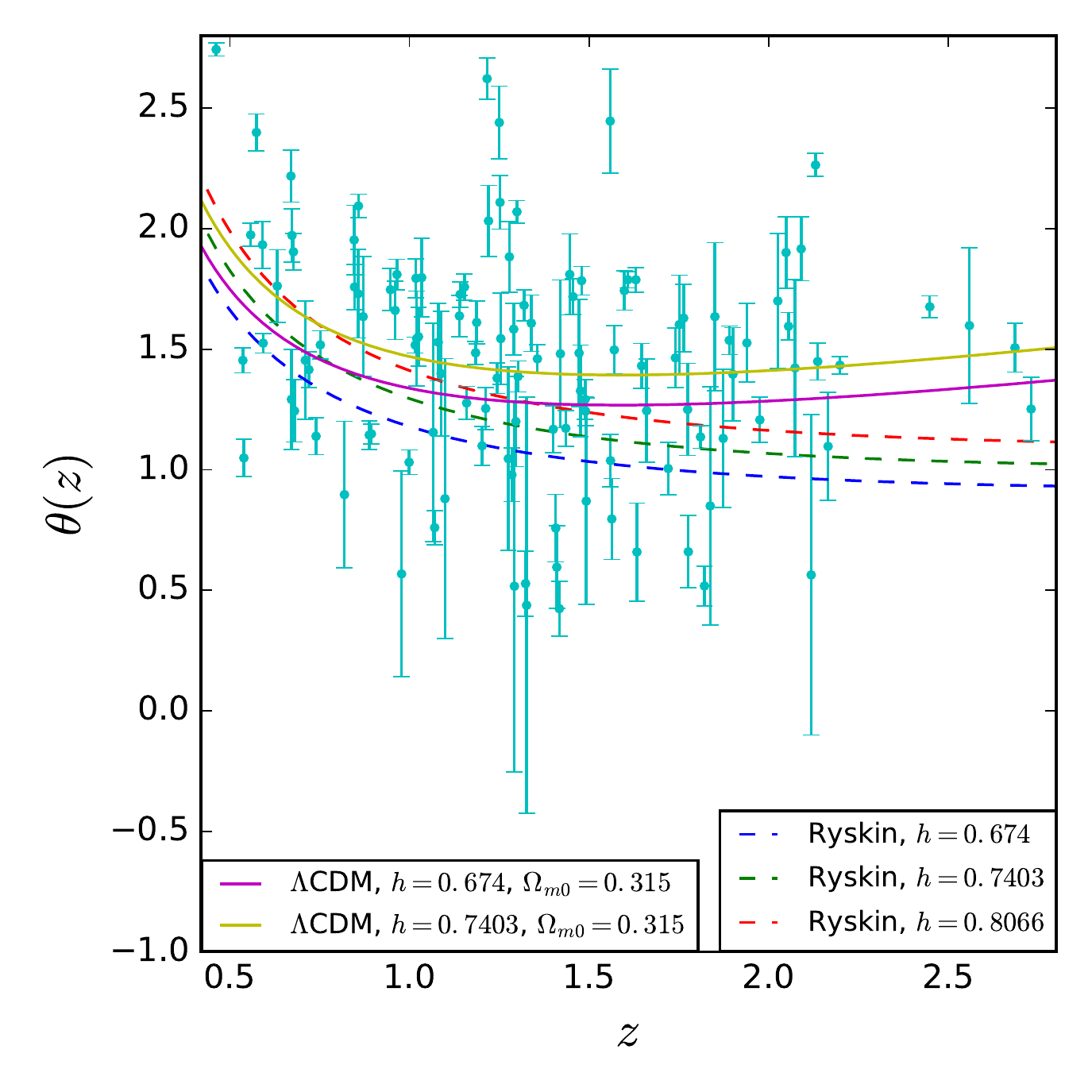}
    \caption{Sound horizon parameter $\theta(z)$ versus redshift $z$ for $\Lambda$CDM and for Ryskin's model. The dashed blue, green, and red curves represent $\theta(z)$ as predicted by Ryskin's model, and the solid purple and gold curves represent $\theta(z)$ as predicted by spatially-flat \lcdm. The values $(h, \om) = (0.674, 0.315)$ come from Ref. \cite{planck2018}, $h = 0.7403$ comes from Ref. \cite{riess2018}, and $h = 0.8066$ comes from the fit of $\theta(z)$ to the QSO data using Ryskin's model. See text for discussion.}
    \label{fig:QSO_Lcomp}
\end{figure*}

\begin{figure*}
    \includegraphics[scale=1]{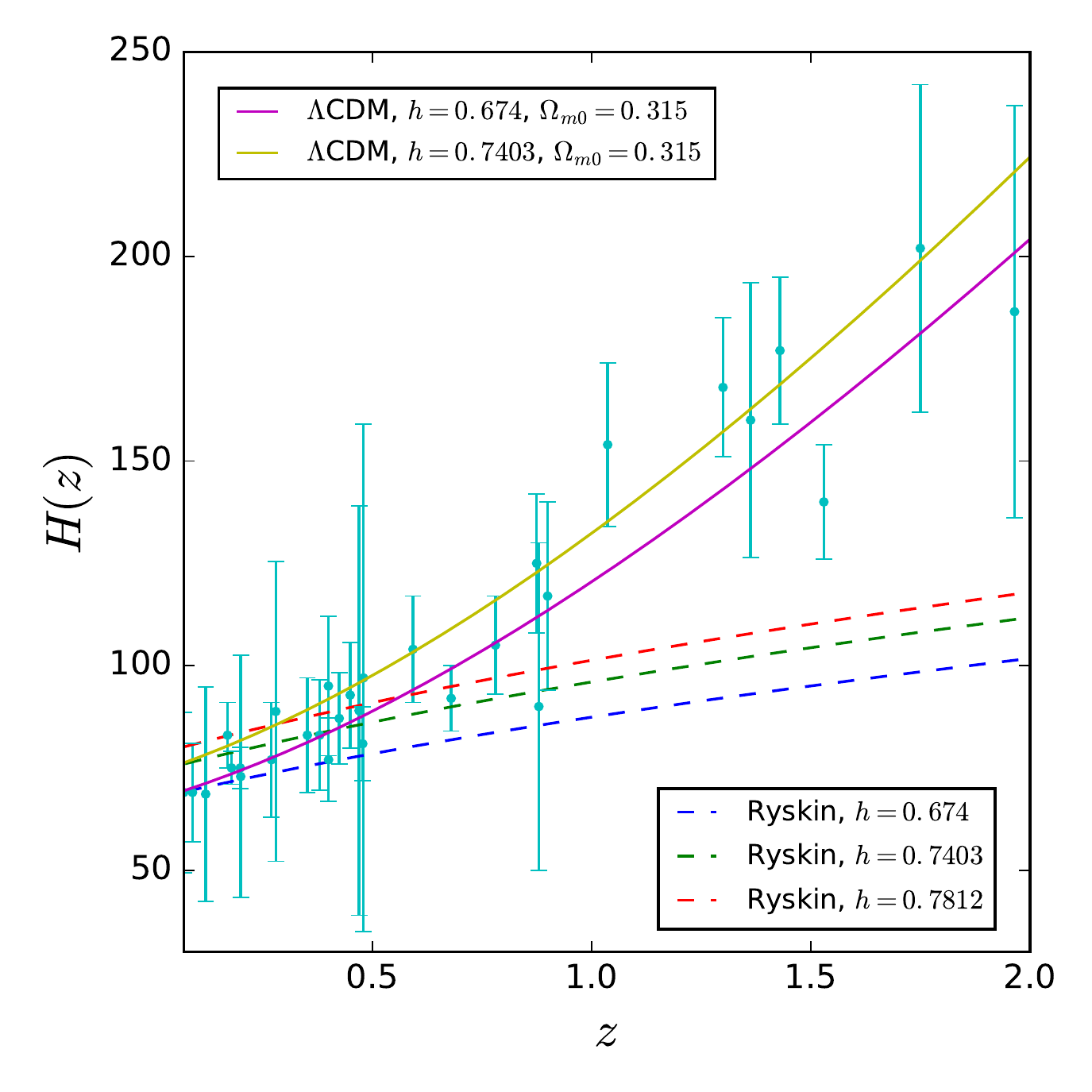}
    \caption{Hubble parameter $H(z)$ versus redshift $z$ for $\Lambda$CDM and for Ryskin's model. The dashed blue, green, and red curves represent $H(z)$ as predicted by Ryskin's model, and the solid purple and gold curves represent $H(z)$ as predicted by spatially-flat \lcdm. The values $(h, \om) = (0.674, 0.315)$ come from Ref. \cite{planck2018}, $h = 0.7403$ comes from Ref. \cite{riess2018}, and $h = 0.7812$ comes from the fit of $H(z)$ to the $H(z)$ data using Ryskin's model. See text for discussion.}
    \label{fig:Hz_vs_z}
\end{figure*}

The results of the comparison between Ryskin's model and the spatially-flat \lcdm\ model using BAO data are presented in Table \ref{tab:BFP_2D_1D_Lcomp} and in Fig. \ref{fig:Lcomp_2D}. In Table \ref{tab:BFP_2D_1D_Lcomp}, the first column lists the model, the second column lists the data set (here ``BAO1" and ``BAO2" have the same meanings as in the previous analysis method), the third and fourth columns list the one-dimensional best-fitting values of $H_0$ and $\Omega_{m0}$, respectively, the fifth column lists the two-dimensional best-fitting values of $H_0$ and $\Omega_{m0}$, and the sixth column lists the value of $\chi^2_{\rm min}/\nu$ corresponding to each model and data set. In Fig. \ref{fig:Lcomp_2D} the left panel shows the two-dimensional constraints on $H_0$ and $\om$ for the BAO1 data combination, and the right panel shows the two-dimensional constraints on $H_0$ and $\om$ for the BAO2 data combination. In both panels the solid contours correspond to Ryskin's model and the dashed contours correspond to \lcdm. From the figure, one can see that even when \lcdm\ and Ryskin's model are both fitted to the low-redshift BAO data (namely BAO2), the confidence contours for both of these models are disjoint to 3$\sigma$. When \lcdm\ and Ryskin's model are fitted to a data combination that includes high-redshift BAO data (namely BAO1), the confidence contours for both models are even more disjoint than when these models are fitted to the low-redshift data.
Fig. \ref{fig:Lcomp_2D} therefore indicates that an investigator who wishes to save Ryskin's model from the BAO measurements I have used must fall on the horns of a dilemma: while it is possible to accommodate a smaller value of $H_0$ within Ryskin's model (i.e. one that is in better agreement with the measurements made by either \cite{planck2018} or \cite{riess2018} and with the \lcdm\ constraints computed here), this is only possible at the cost of predicting an implausibly small value of $\Omega_{m0}$. Similarly, a predicted value of $\Omega_{m0}$ within Ryskin's model that is consistent with the value of $\Omega_{m0}$ predicted by \lcdm\ requires an implausibly large predicted value of $H_0$ within Ryskin's model. Finally, it is clear from Table \ref{tab:BFP_2D_1D_Lcomp} that Ryskin's model, when it is fitted to either BAO1 or BAO2, has a much larger value of $\chi^2_{\rm min}/\nu$ than the \lcdm\ model, when the \lcdm\ model is fitted to the same data combinations. Ryskin's model therefore provides a much poorer fit to the BAO data (especially high-redshift BAO data) than does the standard \lcdm\ model.

In Fig. \ref{fig:QSO_Lcomp} I have plotted $\theta(z)$ vs $z$ for the values of $\theta(z)$ predicted by both Ryskin's model and spatially-flat \lcdm. The dashed blue, green, and red curves represent Ryskin's predicted $\theta(z)$ with $h$ set to 0.674, 0.7403, and 0.7812, respectively. The solid purple and gold curves represent the theoretical curves of $\theta(z)$ as calculated in the spatially-flat \lcdm\ model with ($h$, $\om$) set to (0.674, 0.315) and (0.7403, 0.315), respectively. This plot is, in my view, rather inconclusive with regard to whether spatially-flat \lcdm\ or Ryskin's model provides a better fit to the data, as the measurements are dispersed widely on the plot, and many of them have very large error bars, so the overall trend in the data is difficult to see. It is clear from the theoretical curves, however, that Ryskin's model predicts a very different angular size than \lcdm\ for $z \gtrsim 1.5$, so a stronger case against Ryskin's model from QSO data could potentially be made with more high-redshift measurements (or more precise low-redshift measurements).

The \lcdm\ model departs even more radically from Ryskin's model, at high redshift, when their respective theoretical $H(z)$ curves are plotted against $H(z)$ data. In Fig. \ref{fig:Hz_vs_z} the dashed curves represent Ryskin's predicted $H(z)$, with the blue, green, and red curves corresponding to $h = 0.674$, $h = 0.7403$, and $h = 0.7812$, respectively. The solid curves represent $H(z)$ as predicted by the \lcdm\ model, where the purple curve corresponds to $(h, \om) = (0.674, 0.315)$, and the gold curve corresponds to $(h, \om) = 0.7403, 0.315)$. From the figure, one can see that at low redshift ($z \lesssim 1$), Ryskin's model appears to fit the data as well as \lcdm, owing to the large error bars on the measurements. At high redshift ($z \gtrsim 1$), the data clearly diverge from the curves predicted by Ryskin's model, and the \lcdm\ curves match the upward trend. 
The \lcdm\ model therefore provides a much better fit to high-redshift $H(z)$ data than does Rykin's model.

\section{Conclusion}

I conclude, based on these results and the earlier findings of Ref. \cite{Revisit_Ryskin}, that Ryskin's model of emergent cosmic acceleration does not provide an adequate fit to available cosmological data, and so can not replace the standard spatially-flat \lcdm\ cosmological model. The fit to the SNe Ia data presented in Ryskin's original paper is primarily a fit to low-redshift ($z \lesssim 1$) measurements; as can be seen from Figs. \ref{fig:Lcomp_2D} and \ref{fig:Hz_vs_z}, as well as Tables \ref{tab:BFP} and \ref{tab:BFP_2D_1D_Lcomp}, low-redshift measurements do not clearly distinguish between Ryskin's model and the \lcdm\ model when the predictions of these models are compared to the data. High-redshift measurements, on the other hand (chiefly $H(z)$ and BAO measurements at $z \gtrsim 1$) can distinguish between these two models, and the high-redshift data clearly favor \lcdm\ over Ryskin's model.

\section{Acknowledgments}

Some of the computing for this project was performed on the Beocat Research Cluster at Kansas State University, which is funded in part by NSF grants CNS-1006860, EPS-1006860, EPS-0919443, ACI-1440548, CHE-1726332, and NIH P20GM113109. This work was partially funded by DOE grant DE-SC0019038. I thank Gregory Ryskin for bringing his work to my attention, and I thank Bharat Ratra and the anonymous referee for their helpful comments on drafts of this paper. 

\bibliographystyle{plain}
\bibliography{bibliography2}

\end{document}